\documentclass[a4paper,11pt]{article}
\usepackage{pos}
\usepackage{bm}
\usepackage{graphicx}%

\newcommand{\be}{\begin{equation}}
\newcommand{\ee}{\end{equation}}
\newcommand{\bea}{\begin{eqnarray}}
\newcommand{\eea}{\end{eqnarray}}
\newcommand{\lan}{\langle}
\newcommand{\ran}{\rangle}

\title{ Interactions of $\omega$ mesons in nuclear matter and with nuclei}

\author*[a]{Horst Lenske}

\affiliation[a]{Institut f\"ur Theoretische Physik, Justus-Liebig-Universit\"at Giessen,\\
  Heinrich-Buff-Ring 16, D-35392 Giessen, Germany}

\emailAdd{horst.lenske@physik.uni-giessen.de}

\abstract{In-medium interactions of $\omega$-mesons in infinite nuclear matter and finite nuclei are investigated in a microscopic approach with $N^{-1}$ and $N^*N^{-1}$ particle-hole polarization modes including $N^*$ resonances. The nuclear polarization tensor formalism is used. Covariant mean-field self-energies are taken into account. Longitudinal and transversal self-energies are derived. The theoretical results is applied to finite nuclei in local density approximation.  Applications to recent data on the in–medium width of $\omega$ mesons in $^{93}Nb$ are presented. The data are well described by  self–energies containing S-wave and P-wave $N*$ resonances. Low-energy parameters are determined. For the first time a spectrum of omega-nucleus bound states is obtained, albeit with reduced lifetimes. }

\FullConference{10th International Conference on Quarks and Nuclear Physics (QNP2024)\\
8-12 July, 2024\\
Barcelona, Spain\\}

\begin{document}
\maketitle

\section{Introduction}

Meson self--energies have been studied intensively for many years. The best and longest studied case are interactions of pions with cold and hot nuclear matter, resulting in understanding of the important role of in-medium modes, see e.g. \cite{Oset:1981ih}. In more recent years, the studies were extended to other pseudo-scalar mesons, e.g. to the photo-production of $\eta$-mesons on nuclei \cite{Peters:1998hm} and to $\eta'$-nucleus interactions and search for bound states \cite{Super-FRS:2015etd}. Interactions of vector mesons with nuclei became of special interest signals from highly compressed hot matter \cite{Peters:1997va,Rapp:1999ej}.

Theoretically, in--medium properties and interactions of vector mesons were studied under various aspects from QCD sum rules
\cite{Margvelashvili:1987tk,Braun:1988qv,Hatsuda:1991ez} to spontaneous chiral symmetry breaking \cite{Harada:2016uca}, complemented by a broad
spectrum of phenomenological approaches. Theoretical investigations  of vector meson in--medium spectral functions have  a long tradition as e.g. in
\cite{Celenza:1991ff,Caillon:1995ci,Peters:1997va,Muehlich:2006nn,Shao:2009zzb,Ramos:2013mda,Cabrera:2013zga,Das:2019vhr}, not to the least motivated by the search
for signals of chiral symmetry restoration \cite{Rapp:2009yu}.

In this contribution, interactions of omega-mesons in asymmetric infinite matter and finite nuclei are investigated, following closely the nucleon-nucleon ($NN^{-1}$) and resonance-nucleon ($N^*N^{-1}$) particle-hole  approach discussed in detail recently in \cite{Lenske:2023mis}. Since the omega-meson is an isoscalar particle the coupling is restricted
to isoscalar polarization modes which especially implies that only resonances of isospin $I=\frac{1}{2}$ are allowed. $\omega-A$ interaction studies are especially a filter for in-medium modes involving P-wave and S-wave resonances of type $P_{11},P_{13}$  and $S_{11}$, respectively. Resonances in higher partial waves are increasingly suppressen because of the more pronounced centrifugal barriers.

In the following $\omega$ self-energies from the coupling to polarization modes are investigated. The global features are studied in infinite asymmetric nuclear matter. The results are applied to omega interactions with finite. Advantage is taken of photo-production data for $\omega +{}^{93}Nb$ \cite{Kotulla:2008pjl,Friedrich:2016cms} which were used to extract the energy dependence of the in-medium omega-decay width $\Gamma_{\omega A}$. A systematic bound state search led to a well bound 1s-state, a near-threshold 2s-state and in between a 1p-state. An outlook to further in-medium studies and meson bound states will be given.

\section{Interactions of $\omega$ Mesons in Asymmetric Nuclear Matter}
\begin{figure}[h]
\centering
\includegraphics[width=8cm,clip]{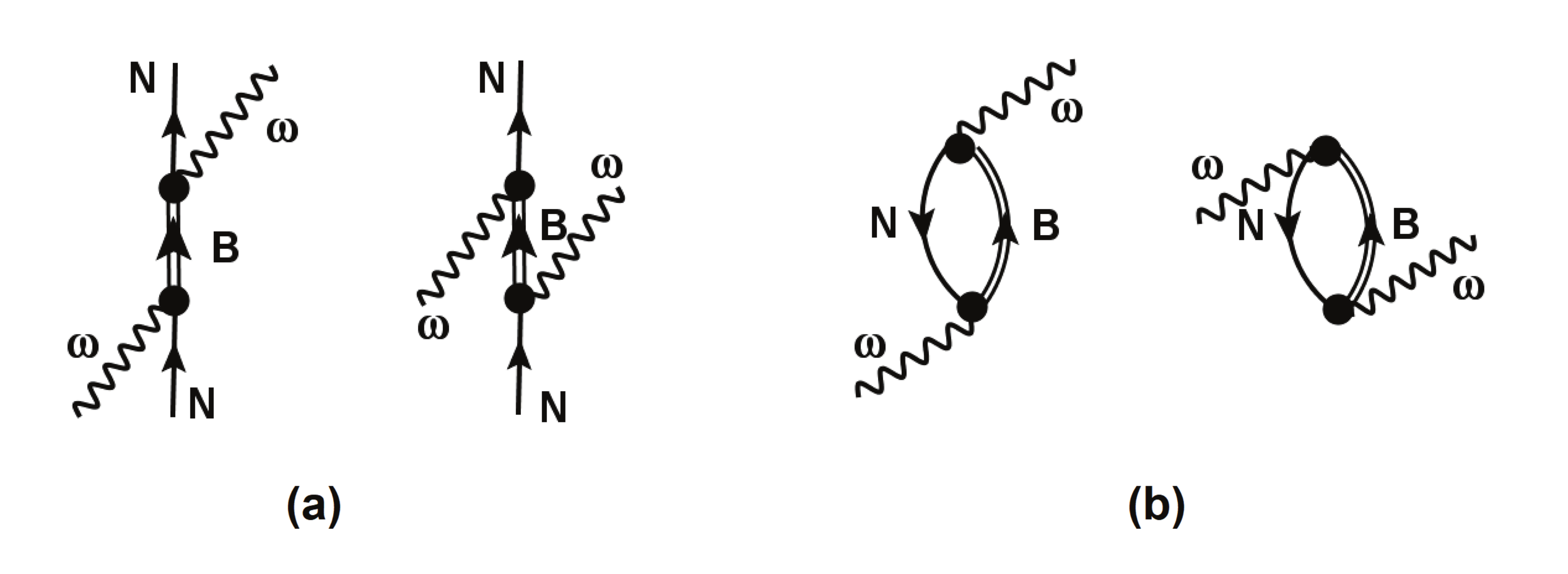}
\caption{The omega--nucleon scattering amplitudes (a) and the corresponding particle--hole type diagrams (b) are depicted. In both cases s--and u--channel diagrams are shown. Nucleons in (a) and hole states in (b) are denoted by $N$. The intermediate $B=N,N^*$ states in (a), which become particle states in (b), are indicated by double lines. }
\label{fig:Graphs}
\end{figure}

As depicted diagrammatically in Fig.\ref{fig:Graphs}, the model is focused on dynamical $\omega+A$ self--energies described by nuclear polarization tensors describing particle--hole excitations of the target
\be\label{eq:PTens}
\Pi^{\mu\nu}_{\alpha\beta}=-tr_A\lan A|\widehat{\Gamma}^\mu_{\beta}\mathcal{G}_A\widehat{\Gamma}^\nu_{\alpha}|A\ran ,
\ee
determined by the ground state expectation value of the transition operators $\widehat{\Gamma}^\nu_\alpha$ and the many--body nuclear Green function $\mathcal{G}_A$. Summations over spin and isospin of the nuclear constituents are indicated by $tr_A$.
The particle--hole type diagrams obtained large attention in early theoretical studies, e.g. \cite{Celenza:1991ff} for the omega--meson and \cite{Peters:1997va} for the rho--meson. In later studies as e.g. \cite{Klingl:1997kf,Lutz:2001mi,Muehlich:2006nn,Rapp:2009yu,Ramos:2013mda} the focus shifted towards in--medium meson loop self--energies, affecting directly the decay channels of the incoming meson. In the following, self--energies from the decay of the omega--meson into other mesons are taken into account on the level of the free space width only. As discussed in \cite{Lenske:2023mis} the (sub--leading) in--medium meson loop diagrams are in fact overlapping partially with the (leading) particle--hole self--energies of Fig. \ref{fig:Graphs}. That leads to an ambiguity for in--medium self--energies with the potential danger of overcounting. Here, the $\omega$ coupling constants to the medium excitations will be determined by fit to the data of Ref.\cite{Friedrich:2016cms}.

The polarization tensors from $BN^{-1}$ modes, $B=N,N^*$, are given by \cite{Lenske:2023mis}
\be
\Pi^{\mu\nu}_{NB}(q)=-i
\int \frac{d^4k}{(2\pi)^4}
Tr_{s,t}\left(\Gamma^{\mu}_{NB} G_N(k|k_{F})\Gamma^{\nu}_{NB} G_B(q+k|k_{F})+(q\leftrightarrow -q)  \right).
\ee
The trace has to be taken over spin and isospin projections.
Fermi--momenta are denoted by $k_{F}=\{k_{F_p},k_{F_n} \}$.  $G_{N,B}$ are in--medium propagators, $N\in \{p,n \}$ and $B\in \{N,N^* \}$. Neglecting rank-2 spin-tensor interactions, only two kinds of isoscalar vertex operators are relevant:
\bea\label{eq:SigmaOp}
&&\Gamma^\mu_{NB}= \gamma^\mu \quad\quad \textrm{if B=N,N$^*$ is a positive parity state}, \\
&&\Gamma^\mu_{NB}= \gamma_5\gamma^\mu \quad \textrm{if B=N$^*$ is a negative parity state}.
\eea
The calculations include  $S_{11}$, $S_{13}$and $P_{11}$, $P_{13}$ resonances, respectively. The self-energy tensor is defined by multiplying the coupling constants to the polarization tensors, summing over the partial contributions, and projections to longitudinal (L) and transversal (T) components are performed:
\be\label{eq:SelfE}
\mathcal{S}^{\mu\nu}_{\omega A}(w,\mathbf{q})=\sum_{N;B}g^2_{\omega NB}\Pi^{\mu\nu}_{NB}(w,\mathbf{q})
=P^{\mu\nu}_{L}\Sigma_L(w,\mathbf{q})+P^{\mu\nu}_{T}\Sigma_T(w,\mathbf{q}),
\ee
where
$\Sigma_{L/T}=\sum_{N=p,n;B=NN^*}g^2_{BN\omega}P^{(NB)}_{L/T}$ is given by the components of the projected polarization tensor.
In our case, the total longitudinal self--energy is fully determined by the longitudinal S--wave component:
\be
\Sigma_L(w,\mathbf{q})\equiv  \Sigma^{(S)}_L(w,\mathbf{q})
\ee
while the total transversal self--energy contains P--wave and S--wave components
\be\label{eq:SigmaT}
\Sigma_T(w,\mathbf{q})= \Sigma^{(S)}_T(w,\mathbf{q})+ \Sigma^{(P)}_T(w,\mathbf{q})
\ee
Accordingly, transversal and longitudinal partial widths are obtained, defined relativistically by the relation:
\be
\Gamma_{L/T}(\mathbf{q})=-\frac{1}{m_\omega}Im\left(\Sigma_{L/T}(E(\mathbf{q}),\mathbf{q})  \right)
\ee
where on--shell kinematics are used. Explicit formulas, details of the numerical calculations, and a list of coupling constants are found in Ref.\cite{Lenske:2023mis}.

\begin{figure}[h]
\centering
\includegraphics[width=8cm,clip]{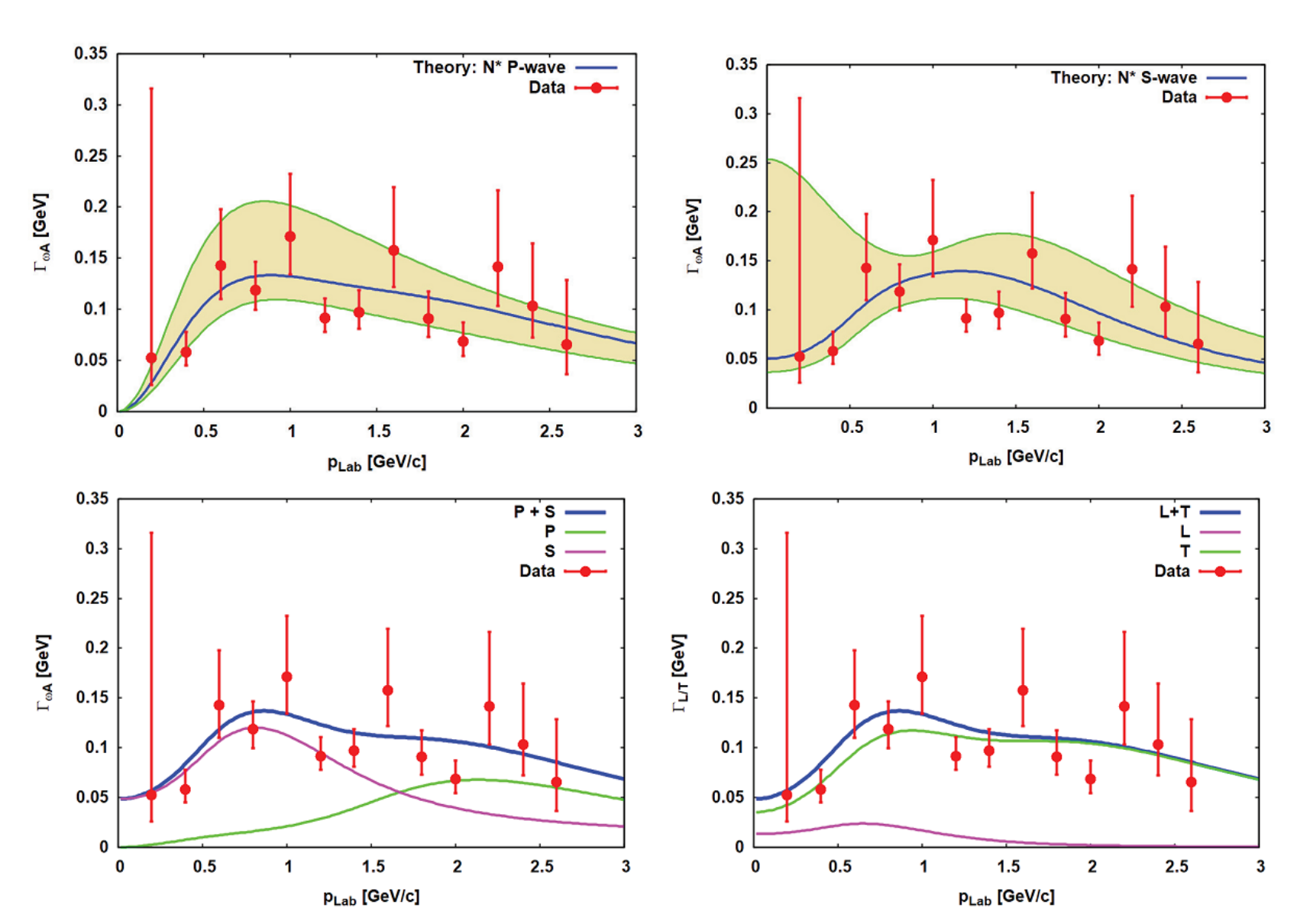}
\caption{The in--medium width of an $\omega$ meson at central density $\rho_A=0.140 fm^{-3}$ of $^{93}$Nb is shown in the top row as function of the omega momentum $p_{Lab}$ and compared to the experimentally deduced values of Ref. \cite{Friedrich:2016cms}.
In the bottom row, P- and S-wave partial widths (left) and the longitudinal and transversal components (right) are displayed.}
\label{fig:Gw_Expl}
\end{figure}

In Fig.\ref{fig:Gw_Expl} the theoretical in-medium widths are compared to the data of Friedrich et al. \cite{Friedrich:2016cms}. In the average the overall description is quite reasonable if only the median value of the calculated width is considered. A look to the error band, however, reveals large uncertainties, mainly due to the large errors of the data in the threshold region. Also shown are the longitudinal and transversal parts and the contributions of the S-wave and P-wave channels.

As discussed in \cite{Lenske:2023mis}, Schr\"odinger-type potentials and low-energy parameters were determined from the full self-energies, including real and imaginary parts. The scattering length and effective range are $a_s=5.6542 -i0.9041$~fm and $r_s=3.7669 -i0.5759~fm$, respectively.

Extrapolation into the nearby below--threshold region allows to explore the bound state as pole of the S--matrix. A bound state was found at energy $\varepsilon_B=-0.488$~MeV and an imaginary part corresponding to a dissipative width $\Gamma^{(A)}_B=4.445$~MeV.

\begin{figure}[h]
\centering
\includegraphics[width=7cm,clip]{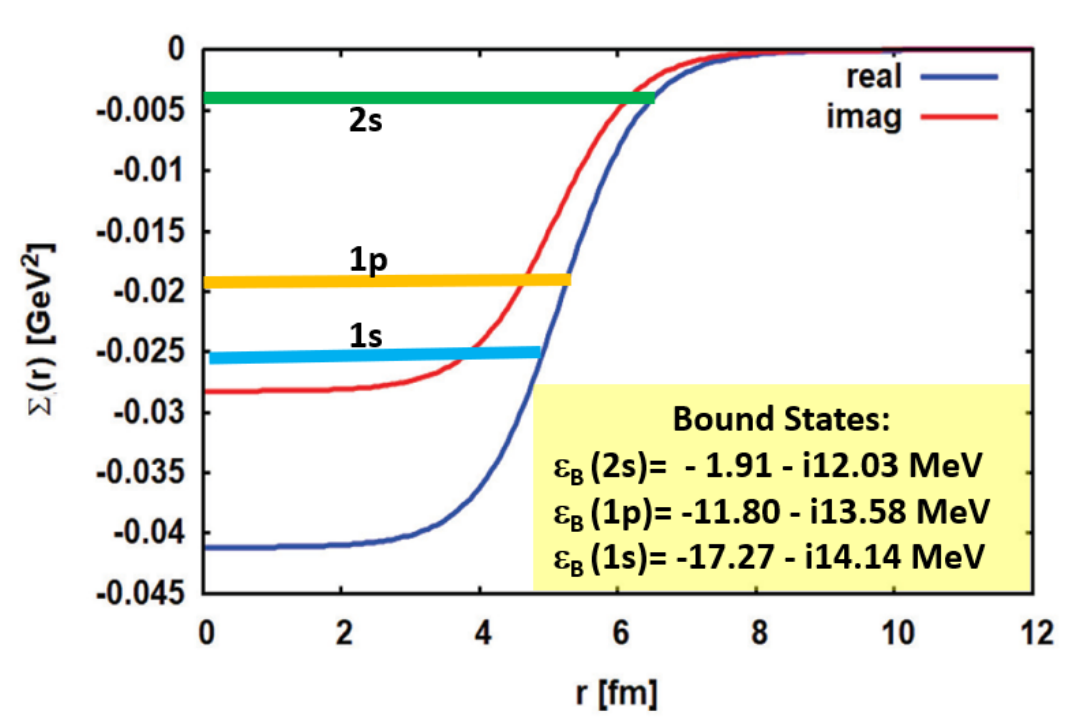}
\caption{
$\omega+{}^{93}Nb$ g.s. self-energies at threshold and the bound state spectrum. The imaginary eigenvalues include the free $\omega$ decay width. }
\label{fig:Bound}
\end{figure}

This result motivates a full search for bound states by solving numerically the $\omega +{}^{93}Nb$ wave equations. The spectrum of $1s$, $2s$, and $1p$  bound states, their complex eigen-energies, and the self-energies at threshold,  respectively, are shown in Fig.\ref{fig:Bound}. The total widths of the states is $\Gamma_{1s}=28.28$~MeV, $\Gamma_{1p}=27.16$~MeV, and $\Gamma_{2s}= 24.06$~MeV reflecting a considerable broadening of the spectral distributions by about a factor of 3, corresponding to a reduction in lifetime by a factor $\sim\frac{1}{3}$. Subtracting $\Gamma_{free}=8.68$~MeV we obtain the in-medium contributions
$\Gamma^{(A)}_{1s}=19.60$~MeV, $\Gamma^{(A)}_{1p}=18.48$~MeV, and $\Gamma^{(A)}_{2s}= 15.38$~MeV, respectively. The
lifetimes of the bound states are about $c\tau_{tot} \sim 5\ldots 6$~fm which is a factor 3 to 4 less than  $c\tau_{free} \approxeq 22$~fm.

\section{Summary and Outlook}
Three important \textit{take-home} messages should be remembered for interactions of omega-mesons with cold nuclear matter:
\begin{itemize}
  \item polarization modes involving nucleon resonances play a central pole for understanding omega in-medium self-energies while purely nucleonic modes play a minor or even negligible role as a background contribution;
  \item precise knowledge of the self-energies at threshold is crucial for fixing coupling constants and controlling accurately the amount of longitudinal and transversal contributions;
  \item for the first time, bound $\omega$-nucleus configurations were obtained.
\end{itemize}
These exploratory investigation underline the high research potential of meson-nucleus physics.
\newline
\paragraph{Acknowledgement:} Inspiring discussions with M. Nanova and V. Metag were of central importance for this project. This work was supported in part by DFG, grants Le439/6 and Le439/7.

\end{document}